\documentclass[preprint,superscriptaddress,aps,prb,noshowpacs]{revtex4-2}
\usepackage{graphicx}
\usepackage{dcolumn}
\usepackage{bm}
\usepackage{color}
\usepackage{amsmath}
\usepackage{epstopdf}
\usepackage{ulem}

\begin{document}

\title{Experimental discovery of Sarma state in atomically thick superconducting FeSe films under high magnetic fields}

\affiliation{State Key Laboratory of Low-Dimensional Quantum Physics, Department of Physics, Tsinghua University, Beijing 100084, China}
\affiliation{Frontier Science Center for Quantum Information, Beijing 100084, China}

\author{Wantong Huang}
\thanks{These authors contributed equally to this work.}
\affiliation{State Key Laboratory of Low-Dimensional Quantum Physics, Department of Physics, Tsinghua University, Beijing 100084, China}

\author{Yuguo Yin}
\thanks{These authors contributed equally to this work.}
\affiliation{State Key Laboratory of Low-Dimensional Quantum Physics, Department of Physics, Tsinghua University, Beijing 100084, China}

\author{Haicheng Lin}
\affiliation{State Key Laboratory of Low-Dimensional Quantum Physics, Department of Physics, Tsinghua University, Beijing 100084, China}

\author{Wei Chen}
\affiliation{State Key Laboratory of Low-Dimensional Quantum Physics, Department of Physics, Tsinghua University, Beijing 100084, China}

\author{Yaowu Liu}
\affiliation{State Key Laboratory of Low-Dimensional Quantum Physics, Department of Physics, Tsinghua University, Beijing 100084, China}

\author{Lichen Ji}
\affiliation{State Key Laboratory of Low-Dimensional Quantum Physics, Department of Physics, Tsinghua University, Beijing 100084, China}

\author{Zichun Zhang}
\affiliation{State Key Laboratory of Low-Dimensional Quantum Physics, Department of Physics, Tsinghua University, Beijing 100084, China}

\author{Xinyu Zhou}
\affiliation{State Key Laboratory of Low-Dimensional Quantum Physics, Department of Physics, Tsinghua University, Beijing 100084, China}

\author{Xusheng Wang}
\affiliation{State Key Laboratory of Low-Dimensional Quantum Physics, Department of Physics, Tsinghua University, Beijing 100084, China}

\author{Xiaopeng Hu}
\affiliation{State Key Laboratory of Low-Dimensional Quantum Physics, Department of Physics, Tsinghua University, Beijing 100084, China}

\author{Yong Xu}
\affiliation{State Key Laboratory of Low-Dimensional Quantum Physics, Department of Physics, Tsinghua University, Beijing 100084, China}
\affiliation{Frontier Science Center for Quantum Information, Beijing 100084, China}
\affiliation{Collaborative Innovation Center of Quantum Matter, Beijing 100084, China}

\author{Lianyi He}
\affiliation{State Key Laboratory of Low-Dimensional Quantum Physics, Department of Physics, Tsinghua University, Beijing 100084, China}

\author{Xi Chen}
\affiliation{State Key Laboratory of Low-Dimensional Quantum Physics, Department of Physics, Tsinghua University, Beijing 100084, China}
\affiliation{Frontier Science Center for Quantum Information, Beijing 100084, China}
\affiliation{Collaborative Innovation Center of Quantum Matter, Beijing 100084, China}

\author{Qi-Kun Xue}
\affiliation{State Key Laboratory of Low-Dimensional Quantum Physics, Department of Physics, Tsinghua University, Beijing 100084, China}
\affiliation{Frontier Science Center for Quantum Information, Beijing 100084, China}
\affiliation{Collaborative Innovation Center of Quantum Matter, Beijing 100084, China}

\author{Shuai-Hua Ji}
\email{shji@mail.tsinghua.edu.cn}
\affiliation{State Key Laboratory of Low-Dimensional Quantum Physics, Department of Physics, Tsinghua University, Beijing 100084, China}
\affiliation{Frontier Science Center for Quantum Information, Beijing 100084, China}
\affiliation{Collaborative Innovation Center of Quantum Matter, Beijing 100084, China}

\date{\today}

\begin{abstract}
Many-body ground states of imbalanced Fermi gas have been studied both theoretically\cite{Sarma63,Takada69,Wilczek03a,Wilczek03b,Badaque03,Wilczek05,Carlson05,Son06,Sheehy06} and experimentally\cite{Hulet06,Zwierlein06a,Hulet06PRL,Shin06,Zwierlein06b,Shin08,Sin08b} for several decades because of their fundamental significance in condensed matter physics\cite{Barzykin07,Zhuang09}, cold atom physics\cite{Chevy06,Hu06,Pao06,Gubbels06,Duan06,Duan06PRA,Parish07,Simons07,Taylor07} and nuclear physics\cite{Bedaque03,Casalbuoni04,Boettcher15}. The Sarma state, a gapless spin-polarized superfluid, is one of those long sought-after exotic ground states of spin imbalanced Fermi gas. Yet, an unambiguous experimental evidence of Sarma superfluid state has not been found\cite{Shin08,Liu19}. Here, we report the experimental discovery of the Sarma state in atomically thick FeSe films by a dilution-refrigerator scanning tunneling microscope under high magnetic fields. In the bilayer or trilayer FeSe films, we directly observe the key evidence of the entrance of the Sarma state: the inner Zeeman splitting coherence peaks cross the Fermi level under high in-plane magnetic fields. The angle dependent critical in-plane magnetic field of coherence peak crossing shows a two-fold symmetry due to the anisotropy of the in-plane g-factor of FeSe films. Moreover, in a superconducting FeSe monolayer of a lateral size of several hundred nanometers, the Sarma state can also be induced by strong out-of-plane magnetic fields. Our findings pave the way to explore the unusual physical properties and potential applications in superconducting spintronics of the spin-polarized Sarma superfluid state.
\end{abstract}

\maketitle

\draft

\vspace{2mm}

The Sarma state is a gapless, spin-polarized superconducting state where the paired and unpaired electrons coexist around the Fermi level and are located in the different regions in momentum space\cite{Sarma63,Giorgini08,Chevy10,Radzihovsky10,Gubbel13}. Initially, the Sarma state is considered to be a thermodynamically unstable ground state in a single band weak-coupling superconductor under a uniform exchange field, which acts only on the spin degree of freedom of electrons\cite{Sarma63}. Ideally, the transition from a BCS superconducting state to a Sarma state requires a magnetic field $B \geq \Delta/\mu_{\rm{B}}$, where $\Delta$ is the superconducting gap and $\mu_{\rm{B}}$ is the Bohr magneton. However, before reaching this critical field, there is a superconductor-to-normal-state transition at a magnetic field $B_{\rm{C}}$ = 0.707$\Delta/\mu_{\rm{B}}$, which is well known as Chandrasekhar-Clogston limit or Pauli limit of superconductors\cite{Chandrasekhar62,Clogston62}. This transition fundamentally prevents the realization of Sarma states in normal superconductors. 


The other obstacle which hinders realizing Sarma states is the orbital pair breaking effect of magnetic fields. Usually in three dimensional bulk materials, the critical field limited by orbital effect is much lower than the one for Sarma states. One possible approach to eliminate the magnetic field orbital effect is to reduce the system to two dimensions and only apply an in-plane magnetic field\cite{Barzykin09}. Therefore, due to those two above factors, the Sarma state is usually a unstable and inaccessible ground state in normal superconductors.

In contrast, recent theoretical studies suggest that Sarma states can be stabilized in a model of momentum-dependent interaction\cite{Wilczek05} or in multiband systems\cite{Barzykin07,Barzykin09,Zhuang09,Subasi10}.  Theoretically, orbital Feshbach resonance could also facilitate the stabilization of Sarma states in cold atom systems\cite{Hu18,Liu19,Hu21}.  In 2006, several cold atom experiments indicate signatures of Sarma states\cite{Hulet06,Zwierlein06a}. However, those states are possibly attributed to the phase-separation state and still under debate\cite{Shin08,Liu19}. So far, the explicit experimental evidences of Sarma states are still absent. 

Here we present unambiguous evidences of the Sarma state of ultra-thin FeSe films under an external magnetic field. In the bilayer or trilayer FeSe films, we apply only an in-plane field which significantly reduces the orbital pair-breaking effect and acts mainly on the spin degree freedom of electrons. By using dilution refrigerator scanning tunnelling microscopy, we directly observe the Fermi level crossing of the inner Zeeman splitting coherence peaks under high in-plane magnetic fields, which is one of the key evidences of the Sarma state. Moreover,  in the superconducting monolayer FeSe, we also achieve a Sarma state by applying a strong out-of-plane magnetic field.

\begin{figure*}[htbp]
  \centering
  \includegraphics[width=6.25in]{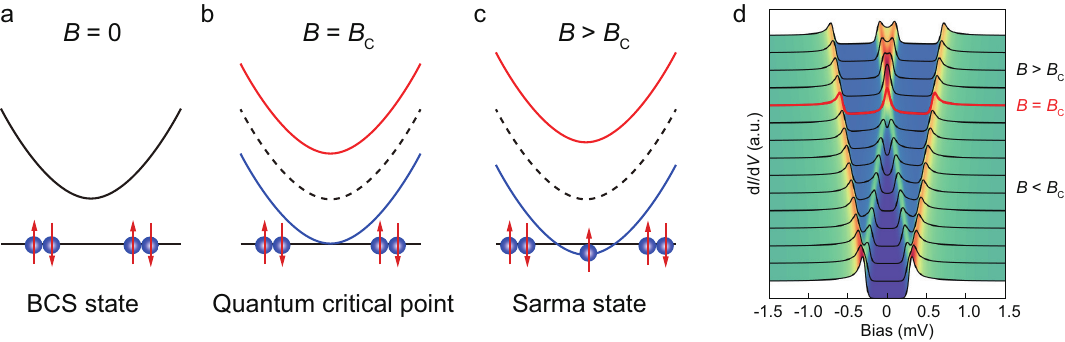}
  \caption{Normal states to Sarma states transition. a, BCS ground state and quasi-particle excitation dispersion curve under zero magnetic field. b, BCS ground state and quasi-particle excitation dispersion curve under the critical magnetic field $B_{\rm{c}}$. The blue and red curves represent two Zeeman splitting branches with spin-up and spin-down polarization, respectively. The bottom of the lower branch of the quasi-particle excitation dispersion reaches BCS ground state. c, the Sarma state with the quasi-particle dispersion lower than the BCS ground state. In momentum space, spin polarized Fermi surface and Cooper pairs are separated. d, the simulated d$I$/d$V$ spectra under a BCS state to a Sarma state transition. The hallmark of this transition is that the inner branches of Zeeman-splitting coherence peak cross the Fermi level. At the critical field, two inner peaks merge at the Fermi level as a single peak. }
\end{figure*}

Our experiments were conducted by using a system which combines molecular beam epitaxy with a dilution-refrigerator scanning tunnelling microscope. The ultra-thin FeSe films of few atomic layers \cite{Xue11,Xue12,Huang21,Lin23} were synthesized by molecular beam epitaxy on graphene layer terminated SiC(0001) substrate. The base temperature of the STM head is 60 mK\cite{Zhao19}, while the effective electronic temperature of our samples is about 228 mK. The Pt-Ir alloy tip of STM was treated by gently contacting a Ag(111) surface until the proper STM topographic images and d$I$/d$V$ spectra were obtained. The in-plane magnetic fields up to 7 T were provided by a split pair superconducting magnet, and all in-plane directions can be accessed by rotating the magnet and/or changing the current direction. The magnetic field direction is calibrated by the in-plane vortex state in a NbSe$_2$ bulk superconductor\cite{Wei11,Jia21}. The monolayer FeSe experiment was conducted under an out-of-plane magnetic field with a magnitude up to 15 T.


The critical transition features from a BCS superconducting state to a gapless Sarma state can be easily identified from the single-particle tunnelling spectroscopy.
First, we discuss a s-wave superconductor with an isotropic superconducting gap. Under a relative small uniform exchange field, BCS ground states are robust, and bogoliubov quasiparticle excitation states exhibit Zeeman splitting(Fig. 1). As the magnitude of the magnetic field increases and the Zeeman splitting energy exceeds 2$\Delta$, the bottom of the lower energy branch of the splitting excited state is lower than the superconducting ground state and a quantum phase transition from a BCS ground state to a Sarma state occurs as shown in Fig. 1(a-c). The Fermi level crossing of the inner Zeeman splitting coherence peaks (Fig. 1(d)) is the hallmark of the entrance of Sarma states. Then, we discuss the case of the superconductors with anisotropic gap, where the gap size is momentum dependent. The critical field to enter Sarma states for the Cooper pairs located at different momentum space would also exhibit momentum-dependent behaviour. The Cooper pairs of smaller gap size usually do not exhibit coherence peaks in $d$I/$d$V spectrum, and it is difficult to identify the entrance of Sarma states for them. However, the coherence peaks of the superconductors of anisotropic gap, which usually exist for the Cooper pairs with larger gap sizes in FeSe films, could also be used to probe the Sarma state. The level crossing of coherence peaks in an anisotropic superconductor is also a hallmark of realizing Sarma states (see Fig. S1 in supplementary materials). 



The Figure 2(a) displays a topographic image of our epitaxial FeSe films. The atomically flat surface and the low defect concentration unveil the high quality of our thin FeSe films\cite{Huang21,Lin23}. The atom-resolved STM image in Fig. 2(b) shows the top Se lattice of the FeSe thin films with the lattice constant of 3.76 {\AA}, which is consistent with bulk FeSe. The local thickness of the FeSe films, which is labelled for different regions in Fig. 2(a), varies from bilayer to quadrilayer. The spatially resolved d$I$/d$V$ spectra on the bilayer FeSe film along the dashed line of 130 nm lateral length in Fig. 2(a) has been shown in the Fig. S2. The superconducting gap size of the middle part is smaller than the one of the both sides (Fig. S2). This is possibly because of the different doping levels from the graphene substrate \cite{Huang21,Lin23} or the proximity effect from nearby trilayer FeSe films. 

The middle superconducting region, where the energy separation between two pronounced coherence peaks is about 0.8 meV, has been chosen for the further study under in-plane magnetic fields because this energy is comparable with the maximum Zeeman splitting energy which can be achieved by our in-plane magnet. The Fig. 2(d) shows the schematic of an in-plane magnetic field applied on the sample. Because the thickness of our FeSe films is two orders of magnitude smaller than the penetration depth of FeSe\cite{Biswas18,Prozorov16,Abdel-Hafiez13}, the magnetic flux could effectively pass into the thin film (Fig. 2(d)). The orbital pairing breaking effect of the in-plane magnetic field on the superconducting state of atomically thin films can be neglected, and the field mainly has the effect on the spin of electrons. It is evidenced by the fact that no vortex state has been observed under in-plane magnetic fields (see Fig. S3 in supplementary materials).

\begin{figure*}[htbp]
  \centering
  \includegraphics[width=6.25in]{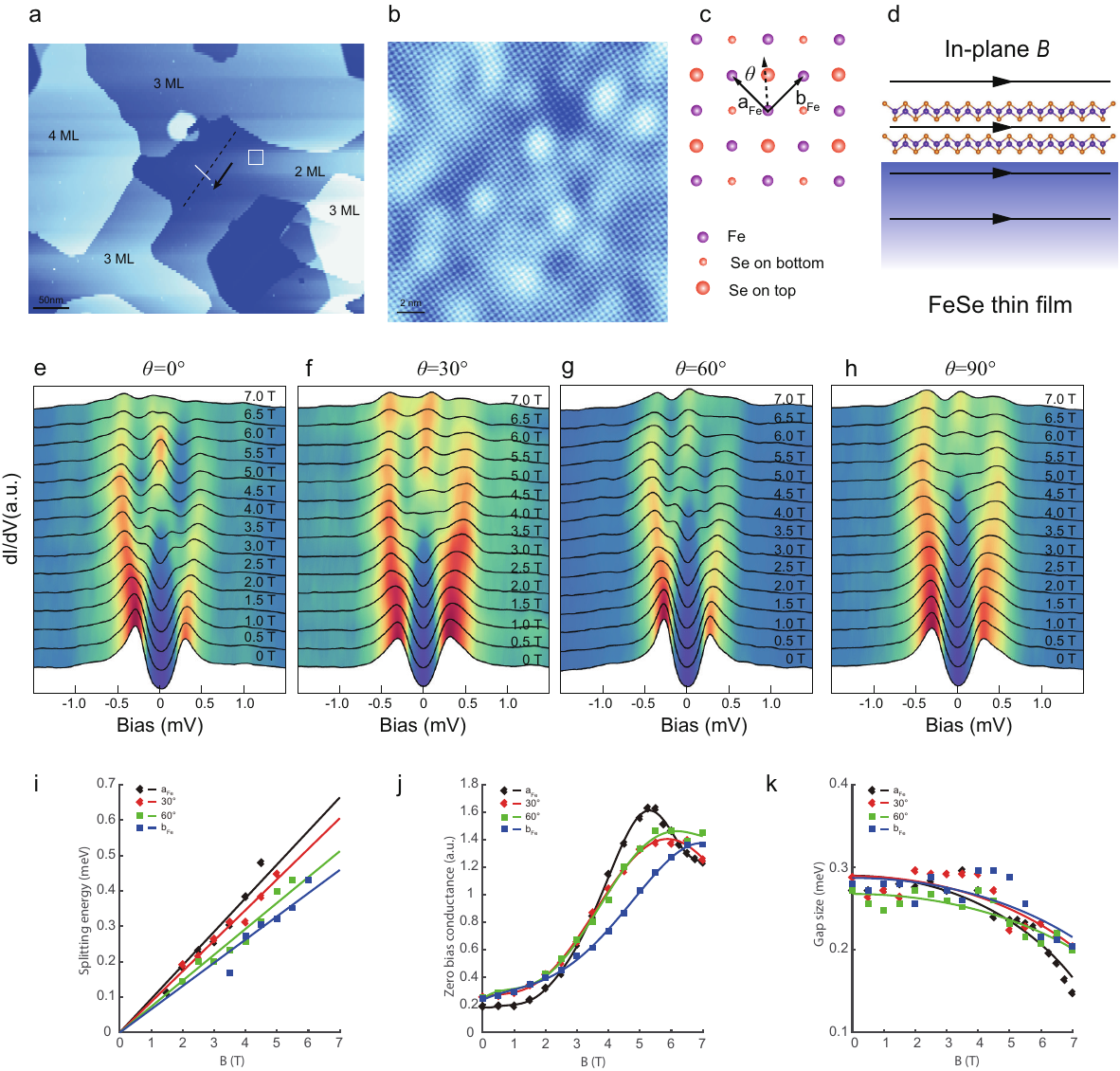}
  \caption{The in-plane magnetic field induced Sarma states in bilayer FeSe films. a, topography image of FeSe films of atomically thick FeSe films (image size: 500 $\times$ 400 nm$^2$, $V_B$ = 3 V, $I_t$ = 20 pA, $T$ = 60 mK). The thickness of FeSe films has been identified as 2 ML, 3 ML and 4 ML, as marked in the each region. b, atomically resolved FeSe film surface ($V_B$ = 0.6 V, $I_t$ = 100 pA, $T$ = 4.95 K). The square lattice of top Se atoms is revealed. c, schematic of top ML FeSe thin film. The solid arrows indicate the primitive vectors. d, the side view of bilayer FeSe under an in-plane magnetic field. e-h, a series of magnetic-field-dependent d$I$/d$V$ spectra with $\theta$ from 0$^\circ$ to 90$^\circ$, $\theta$ = 0$^\circ$ for (e), $\theta$ = 30$^\circ$ for (f), $\theta$ = 60$^\circ$ for (g) and $\theta$ = 90$^\circ$ for (h) ($V_B$ = 3 mV, $I_t$ = 100 pA, $T$ = 60 mK). i-k, angle-dependent Zeeman splitting energy (i), zero bias conductance (j) and gap size (k) as a function of in-plane magnetic fields. }
\end{figure*}

At first, we apply the magnetic field along the a$_{\rm {Fe}}$ direction which is marked in Fig. 2(c) ($\theta$ = 0$^\circ$). As seen in the d$I$/d$V$ spectra of Fig. 2(e), the four sets of superconducting coherence peaks indicate the Zeeman splitting under in-plane magnetic fields up to 7.0 T, and separation of split peaks increases with the increasing magnetic field. Around 5.5 T, the inner two branches of peaks merge at zero bias. The further increased magnetic field leads to the crossing of the coherence peaks over Fermi level, which is the key evidence that a stable Sarma state consisting of spin polarized Fermions and Cooper pairs has been formed. We should point out that the critical field for the entrance of Sarma states for the FeSe bilayer could be much lower than 5.5 T because the Copper pairs with a smaller gap size could enter the Sarma state at a lower magnetic field. However, it is difficult to clearly identify the Sarma state at a low magnetic field as shown in Fig. S1(b) and (c), where the difference between the superconductors with and without Pauli limit is difficult to distinguish.


Next, we change the direction of magnetic fields to $\theta$ = 30$^\circ$, 60$^\circ$ and 90$^\circ$. The d$I$/d$V$ spectra under in-plane magnetic fields up to 7 T are presented in Fig. 2(f-h). The similar Zeeman splitting of superconducting coherence peaks has been observed. However, the critical field for the Fermi level crossing of coherence peaks is enlarged from 5.5 T to a higher magnitude. Figure 2(i) shows the Zeeman splitting energy as a function of an in-plane magnetic field at four different directions. The Zeeman splitting energy exhibits a linear relationship with the magnetic field, whereas the slopes of the linear curves of Fig. 2(i) are slightly reduced from a$_{\rm {Fe}}$ direction to b$_{\rm {Fe}}$ direction indicating the decreased, anisotropic in-plane g factor of the bilayer FeSe film. The curves of zero bias conductance as a function of magnetic field are presented in Fig. 2(j). The peak position of those curves indicates the approximate critical field to realize the Fermi level crossing of coherence peaks. With the increasing $\theta$, the magnitude of the magnetic field corresponding to the maximum zero bias conductance gradually shifts to higher values. The separation between two coherence peaks with the same spin polarization indicates the superconducting gap size (see Fig. S4 in supplementary materials). As the magnetic field increases, the gap size remains almost constant at low fields up to 4 T, but then it considerably decreases as the field reaches 7 T, indicating the suppression of the superconducting gap by the in-plane magnetic field. This observation is consistent with the previous theoretical results calculated at finite temperature\cite{Sarma63}, and the field dependent gap size is fitted by $(\frac{\Delta(T,B)}{\Delta(T,0)})^2=1-(\frac{B}{B_{\rm{C}}})^2$\cite{Douglass61} (Fig. 2(k) and Fig. S6(c) in supplementary materials).

The anisotropy of the critical magnetic field required for the Fermi level crossing of coherence peaks can be explained by the in-plane anisotropy of the effective g-factor. The anisotropy of the g-factor is governed by the spin-orbit interaction and the symmetry of the system. Such anisotropic g-factors have been observed not only in bulk materials\cite{Roth59,Hensel69,Chen85} but also quantum well\cite{Nefyodov11,Eldridge11,Shchepetilnikov15,Minkov20} and quantum dots\cite{Camenzind21} with reduced dimensions. The prominent spin-orbit coupling strength\cite{Coldea15,Borisenko16,Li17} and the inversion symmetry breaking of FeSe films on the graphene surface could lead to the anisotropy of the in-plane g-factor \cite{Eldridge11}. This effect in the bilayer FeSe film is evidenced in Fig. 3(a) and exhibits a two-fold symmetry. Along the a$_{\rm{Fe}}$ axis, the in-plane g-factor reaches the minimum of 1.2 and increases to the maximum of 1.6 along the b$_{\rm{Fe}}$. The anisotropic g-factor can be well fitted by the formula: $g(\theta) = \sqrt{{g_a}^2cos(\theta)^2+{g_b}^2sin(\theta)^2}$, where $\theta$ is the angle between in-plane magnetic field direction and a$_{\rm{Fe}}$ axis, $g_a$ and $g_b$ are the in-plane g-factors along the a$_{\rm{Fe}}$ and b$_{\rm{Fe}}$ axes, respectively. Due to the anisotropy of the in-plane g-factor, the critical field to realize the Fermi level crossing of coherence peaks also exhibits a two-fold symmetry as shown by the dots in Fig. 3(b). The experimental data can be well fitted by the formula
\begin{equation}
B=\frac{2\Delta_0}{\sqrt{(g(\theta)\mu_{\rm{B}})^2+\frac{4\Delta_0^2}{B_{\rm{C}}^2}}}
\end{equation}
, where $\Delta_0$ is the gap size without magnetic field, $g(\theta)$ is the angle dependent g-factor and $B_{\rm{C}}$ is the fitting parameter for field dependent gap size, as shown by solid line (Fig. 3(b)). Along the a$_{\rm{Fe}}$ axis, the critical field is around 7 T because of the small g-factor, while the critical field is reduced to 5.5 T for the magnetic field along the b$_{\rm{Fe}}$ axis due to the larger g-factor.

\begin{figure*}[htbp]
  \centering
  \includegraphics[width=6.25in]{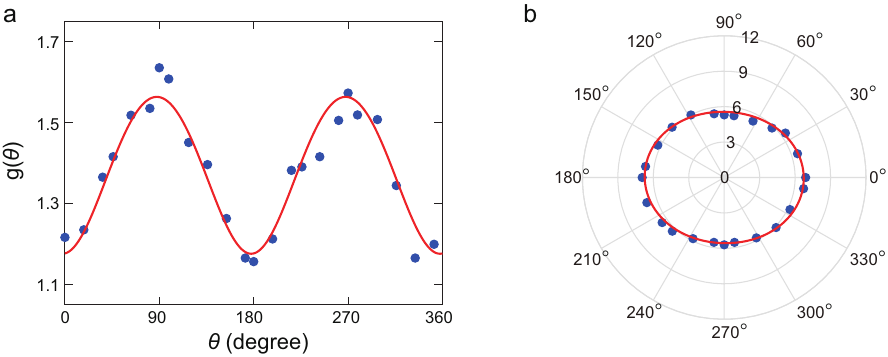}
  \caption{Angle dependent Sarma states. a, angle-dependent in-plane g factors of FeSe bilayer. The blue dots are from the experiment data. The red line is the fit curve from the magnetic-field-dependent Zeeman-splitting energy . b, angle-dependent critical field for the Fermi level crossing of Zeeman splitting coherence peaks. The blue dots are determined from ZBC peaks in the Fig. 2(j). The red line is the fit curve using formula 1. }
\end{figure*}

For a Sarma state, one part of the Fermi surface is gapped and the other part is in the normal metal state in momentum space.  However, in real space, the spectrum of the Sarma state should be uniform. That means the crossing Fermi energy level of Zeeman splitting coherence peaks should be uniform in real-space. The spatially resolved d$I$/d$V$ spectra are shown in Fig. 4(a). Under the zero magnetic field, the relatively uniform coherence peaks with small peak position fluctuations as seen in Fig. 4(a) are observed along the white line in Fig. 2(a). As the magnetic field increases to 3 T, the Zeeman splitting of coherence peaks can be resolved. The splitting of coherence peaks is much clearer in the second derivative of d$I$/d$V$ spectra (Fig. 4(b)). When the magnetic field reaches 6 T, the two inner splitting coherence peaks meet at the Fermi level resulting in a pronounced peak at the zero bias voltage. The further increased magnetic field leads to the Fermi level crossing of coherence peaks, exhibiting the evidence of the Sarma state.

\begin{figure*}[htbp]
  \centering
  \includegraphics[width=6.25in]{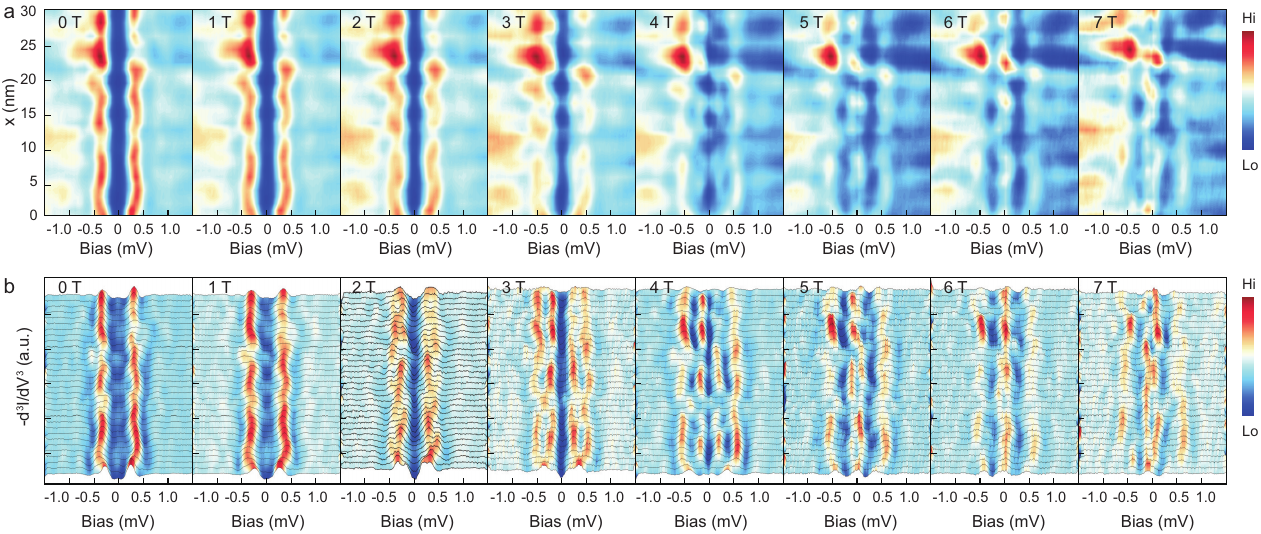}
  \caption{Spatial distribution of Sarma states. a, the spatially resolved d$I$/d$V$ spectra along a dashed line (30 nm) in Fig. 2a at magnetic fields from 0 T to 7 T ($V_B$ = 3 mV, $I_t$ = 100 pA, $V_{\rm{osc}}$ = 0.03 mV, $T$ = 60 mK). b, the spatially resolved of -d$^3I$/d$V^3$ derived from a.}
\end{figure*}

In addition to the bilayer FeSe film, the Sarma state also has been observed in a trilayer region which is evidenced in Fig. S5 and S6 of the supplementary materials. Moreover, we observe this Sarma state as well in monolayer FeSe films under an out-of-plane magnetic field. In the monolayer FeSe film with lower superfuid density, the out-of-plane magnetic field could effectively penetrate into the superconducting film and induce the Zeeman splitting of the coherence peaks. As shown Fig. 5(a), the d$I$/d$V$ spectra of the monolayer FeSe film (the topography image, spectrum location and the spatially resolved d$I$/d$V$ spectra of this region are shown in Fig. S7) under perpendicular magnetic fields up to 15 T reveal the Zeeman splitting of coherence peaks. As the magnetic field exceeds 9 T, the zero bias conductance reaches a maximum and then decreases (another example on monolayer FeSe film is shown in Fig. S8). This behavior closely resembles that observed in bilayer FeSe under an in-plane magnetic field. We believe that Sarma states can be realized in a monolayer FeSe films under a large out-of-plane magnetic field. However, for the bilayer or thicker FeSe films, the out-of-plane magnetic field leads to a vortex state instead of a Sarma state.

In the last part, we would like to discuss the mechanism of the stable Sarma states in FeSe thin films. One of the possible mechanisms is the momentum-dependent interaction\cite{Wilczek05} proposed by Forbe and co-authors. The strong anisotropic pairing gap of bulk FeSe has been studied by scanning tunneling spectroscopy and angle-resolved photoemission spectroscopy \cite{Xue11,Hanaguri14,Davis17,Zhou18}. Under zero magnetic field, this momentum-dependent pairing strength of our FeSe films is also evidenced by the V-shaped superconducting gaps as shown in the Fig. 2e and Fig. S5. Another possible mechanism is the inter-band exchange interaction proposed by He and co-authors \cite{Zhuang09} given that FeSe is a natural multi-band superconductor where several electron and hole bands contribute to superconductivity. He and co-authors also suggest that a Sarma state can be stable in some range of the parameters of a two-band model. We should also notice that, because of the anisotropic gap of FeSe films, how the spin-polarized superconducting Sarma state is formed in the whole momentum space is still not clear. The comprehensive understanding of the underlying mechanisms governing the stable Sarma state requires further theoretical and experimental investigations.

\begin{figure*}[htbp]
  \centering
  \includegraphics[width=6.25in]{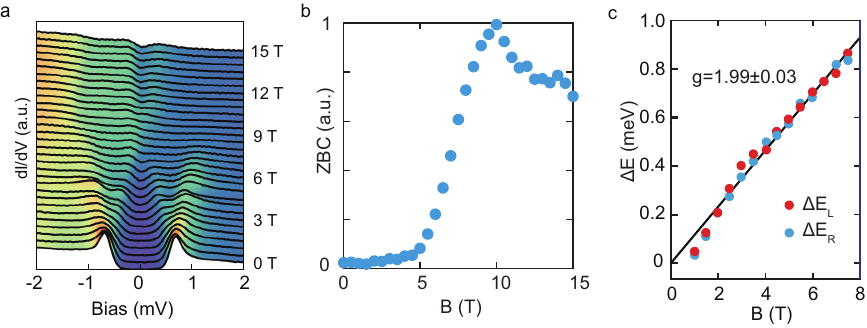}
  \caption{Sarma states of FeSe monolayer under out-of-plane magnetic fields. a, out-of-plane magnetic field-dependent d$I$/d$V$ spectra of FeSe monolayer measured from 0 T to 15 T ($V_B$ = 5 mV, $I_t$ = 100 pA, $V_{\rm{osc}}$ = 0.05 mV). The spectra  are offset for clarity. b, magnetic field dependence of zero bias conductance, extracted from a. c, Zeeman splitting of coherence peaks as a function of magnetic field. $\Delta{\rm{E_L}}$ and $\Delta{\rm{E_R}}$ represent the splitting of the left and right coherence peaks, respectively. A fit yields an out-of-plane g-factor of 
1.99 $\pm$ 0.03.}
\end{figure*}

In summary, the Sarma state has been experimentally discovered in FeSe bilayer and trilayer films under in-plane magnetic fields, as well as in FeSe monolayer films under out-of-plane magnetic fields. We have demonstrated the Fermi level crossing of the inner branches of Zeeman splitting coherence peaks of those atomically thick FeSe films under magnetic fields, which is the unambiguous, hallmark experimental evidence of Sarma states. In addition, due to the in-plane g-factor anisotropy, we have observed an anisotropic effect on the in-plane critical magnetic field for the Fermi level crossing of the Zeeman splitting coherence peaks in bilayer or trilayer FeSe films. Our work unravels the exotic gapless superconducting state with the occupied spin-polarized Bogoliubov quasiparticles, which may lead to unusual paramagnetic Meissner effect\cite{Zhuang06}. The discovery of the Sarma state in solids paves the way to further study the transport\cite{Zwierlein11} and other exotic properties, and may lead to potential electronic applications in spintronics. Although the Sarma states in our experiment are observed in two-dimensional thin films, they would also be possibly realized in three dimensional bulk materials with the intercalation strategy \cite{Chen18,Zhou22} or in two-dimensional-like bulk states with layered structures.

\begin{acknowledgments}
This work was supported by the National Natural Science Foundation of China (Grant No. 12074211, No. 12141403, No. 52388201), and the Ministry of Science and Technology of China (Grants No.2023ZD0300500).
\end{acknowledgments}


%

\clearpage

\noindent\textbf{Supplementary material}

\renewcommand\thefigure{S\arabic{figure}}
\setcounter{figure}{0}

\section{Anisotropic superconductor}
\begin{figure}[htp]
	\begin{center}
		\includegraphics[width=5in]{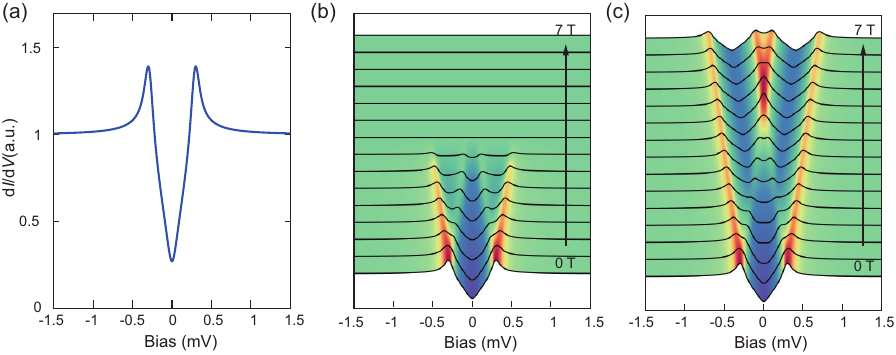}
	\end{center}
	\caption[]{\label{AnisotropicSC}(a) The simulated d$I$/d$V$ spectrum of an anisotropic superconductor with a d-wave like superconducting gap. ($T$ = 260 mK, $\Delta(\theta)=\Delta_0\rm{cos}(\theta)^2$, $\Delta_0$ = 0.29 meV.) (b-c) The simulated d$I$/d$V$ spectrum of an anisotropic superconductor under in-plane magnetic fields with (b) and without (c) Pauli limitation. The level crossing at the Fermi level of coherence peaks indicates the realization of the Sarma state in an anisotropic superconductor.
	}
\end{figure}

FeSe exhibits strongly anisotropic superconducting gap structure. Here we present the simulated d$I$/d$V$ spectrum of FeSe superconducting gap with a d-wave like gap structure as shown in Fig. \ref{AnisotropicSC}(a). It clearly shows the V-shaped gap near the Fermi level and pronounced coherence peak at bias $V$ = $\pm\Delta_{\rm{max}}$. If FeSe belongs to the superconductor with Pauli limitation, the coherence peak would not cross Fermi level before the superconductivity is completely suppressed under in-plane magnetic fields as shown in Fig. \ref{AnisotropicSC}(b). However, the realization of Sarma states in FeSe film can be identified as the coherence peaks cross Fermi level under in-plane magnetic fields (Fig. \ref{AnisotropicSC}(c)). Here we clearly show that coherence peaks of anisotropic superconductors can be used to probe the Sarma state. Under the low magnetic fields from 0 to 4 T, the difference between superconductors with and without Pauli limit is difficult to be distinguished.

\clearpage

\section{The spatially resolved d$I$/d$V$ spectra of the bilayer FeSe film}

\begin{figure}[htp]
	\begin{center}
		\includegraphics[width=5in]{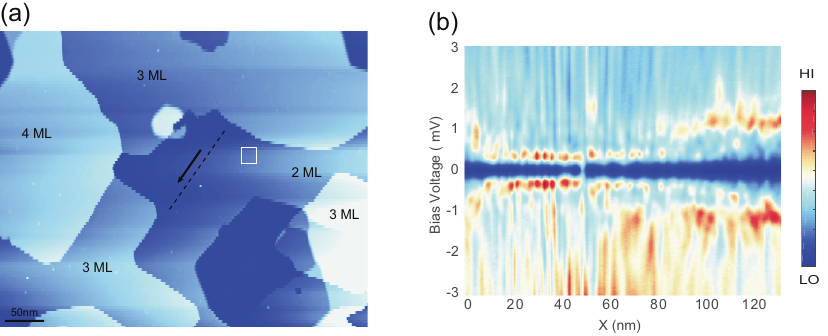}
	\end{center}
	\caption[]{\label{LineSpectra}(a) Topographic image (460 nm $\times$ 370 nm) of FeSe film acquired by using sample bias of $V$=3V and tunneling current of $I$=20 pA. The thickness of FeSe films from bilayer to quadrilayer is labelled for each region by 2 ML, 3 ML and 4 ML. (b) A series of d$I$/d$V$ spectra along a black arrow of 130 nm long in (a) (128 points evenly distributed along this dashed line). Set point: $V_{\text{s}}$ = 3 mV, $I_{\text{t}}$= 100 pA, $V_{\text{osc}}$= 0.3 mV.
	}
\end{figure}

The topographic image of FeSe films is shown in Fig. \ref{LineSpectra}(a), and the thickness of these films are from bilayer up to quadrilayer. A series of d$I$/d$V$ spectra (Fig. \ref{LineSpectra}(b)) taken along the black dashed line of 130 nm long in Fig. \ref{LineSpectra}(a) exhibit strong inhomogeneity of superconductivity. The shape and size of the superconducting gap show a large variation in different regions. The gap size $\Delta$ of the middle part is smaller than the one of both sides. 

\clearpage

\section{The d$I$/d$V$ maps of the bilayer FeSe films under an in-plane magnetic field}
The superconducting coherent length along c axis of bulk FeSe is about 2.2 nm which is larger than the thickness of bilayer film. Therefore, vortexes can not be induced by an in-plane magnetic field in the bilayer FeSe films. In order to rule out the influence of vortex, d$I$/d$V$ maps at zero bias voltage under different in-plane magnetic field were measured. The contrast of zero bias conductance maps originates from inhomogeneous superconductivity, and no vortex has been observed. This demonstrates that the in-plane magnetic fields are nearly perfectly aligned with the sample surface.

\begin{figure}[htp]
	\begin{center}
		\includegraphics[width=4in]{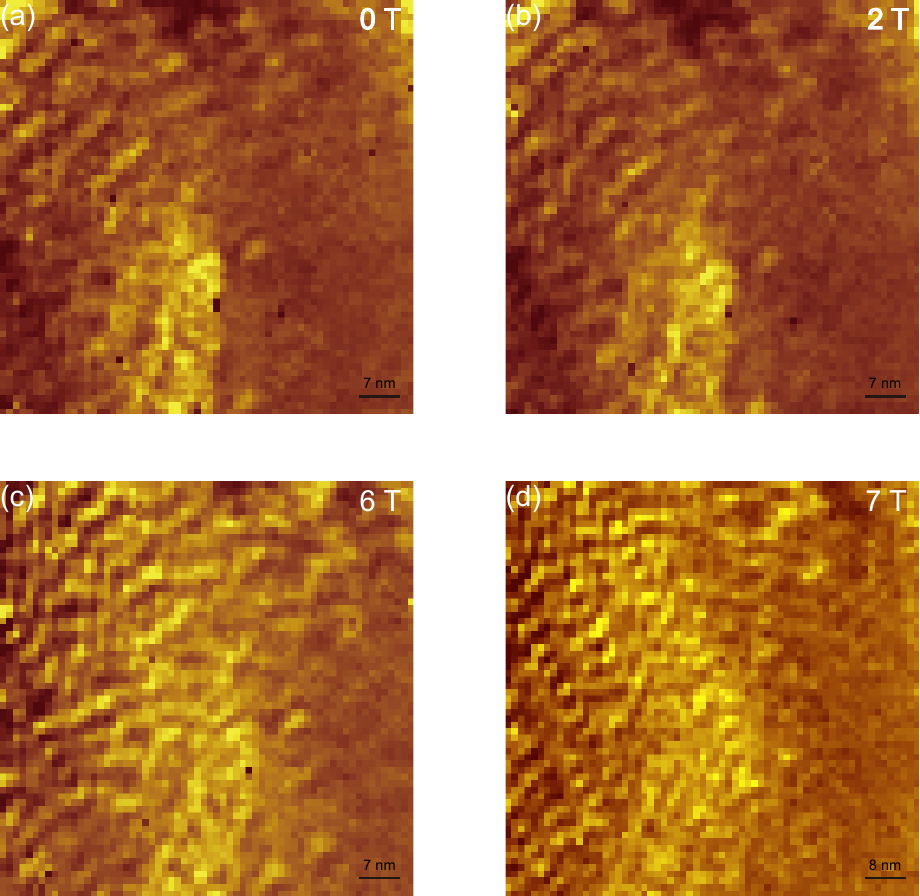}
	\end{center}
	\caption[]{\label{dIdVMap}(a-d) The d$I$/d$V$ maps $L(\textbf{r}, E)$ at zero bias voltage under 0 T, 2 T , 6 T and 7 T in-plane magnetic fields. (a-b) The maps were taken on a grid of 64 $\times$ 64 pixels for a 70 nm $\times$ 70 nm field of view. Sample bias voltage $V_{\rm{s}}$ = 30 mV, tunneling current $I_{\rm{t}}$ = 100 pA, modulation amplitude for the lock-in detection $V_{\text{osc}}$ = 0.3 mV. (c) 64 $\times$ 64 pixels for a 70 nm $\times$ 70 nm field of view. $V_{\rm{s}}$ = 20 mV,  $I_{\rm{t}}$ = 100 pA,  $V_{\text{osc}}$ = 0.2 mV. (d) 64 $\times$ 64 pixels for a 80 nm $\times$ 80 nm field of view. $V_{\rm{s}}$ = 20 mV, $I_{\rm{t}}$ = 100 pA, $V_{\text{osc}}$ = 0.2 mV.
	} 
\end{figure}
\clearpage

\section{Spin polarized Zeeman splitting coherence peaks}
\begin{figure}[htp]
	\begin{center}
		\includegraphics[width=3.5in]{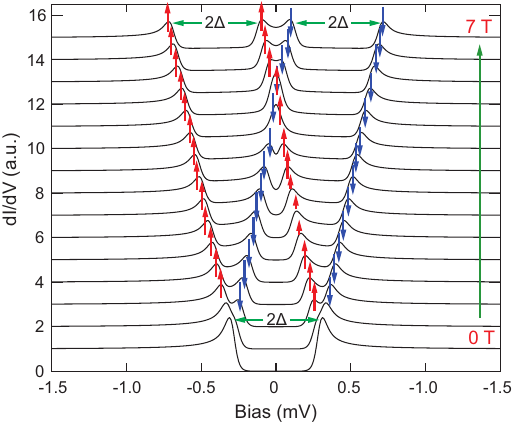}
	\end{center}
	\caption[]{\label{SpinSplit}The simulated d$I$/d$V$ spectra of 2D superconductor under in-plane magnetic fields. It shows the Zeeman splitting of superconducting coherence peaks. The spin-polarization of each peak is indicated by red and blue arrows.
	}
\end{figure}

The Bogoliubov quasi-particle state would show a Zeeman splitting when an in-plane magnetic field (the field only acts on the spin degree of freedom of electrons) is applied as shown in Fig. 1. Hence, the superconducting coherence peaks would exhibit a Zeeman splitting under in-plane magnetic fields. Each coherence peak in Fig. \ref{SpinSplit} is spin-polarized and the polarization direction of each peak is marked by a nearby red (spin-up state) or blue (spin-down state) arrow. Without magnetic field, the energy separation between coherence peaks is 2$\Delta$, and the energy separation between peaks with same spin direction is also 2$\Delta$ which indicates superconducting gap size. Here, we assume the gap size is independent of magnetic fields.

\clearpage

\section{Sarma states of the trilayer FeSe film under in-plane magnetic field}
\begin{figure}[htp]
	\begin{center}
		\includegraphics[width=5in]{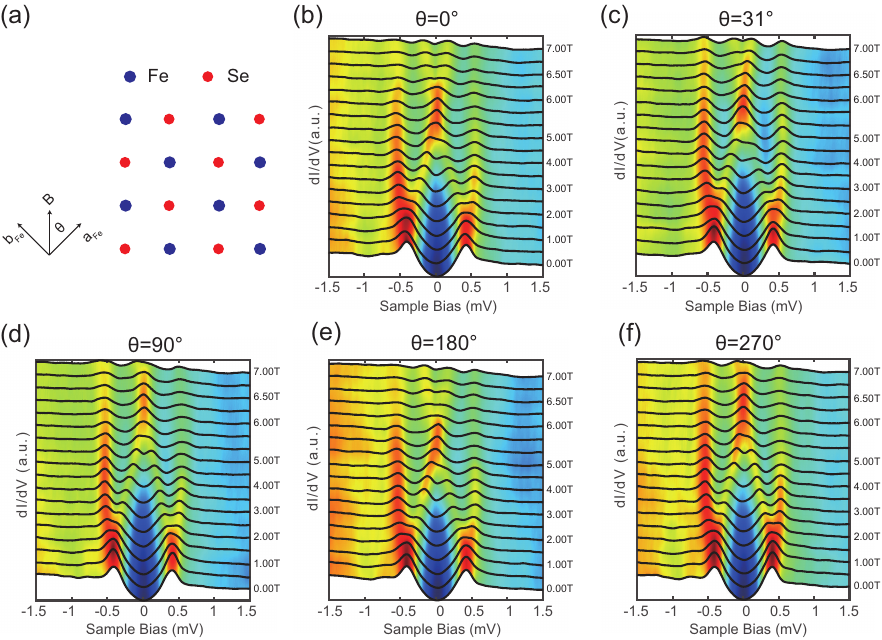}
	\end{center}
	\caption[]{\label{TrilayerdIdV}(a) The top view of FeSe layer lattice structure. The red solid circles are Fe atoms and blue solid circles are Se atoms. (b-f) the d$I$/d$V$ spectra of trilayer FeSe film under in-plane magnetic fields of different directions. $\theta$ is the angle between the field and a$_{\text{Fe}}$. (b) $\theta$ = 0$^{\circ}$, (c) $\theta$ = 31$^{\circ}$, (d) $\theta$ = 90$^{\circ}$, (e) $\theta$ = 180$^{\circ}$, (f) $\theta$ = 270$^{\circ}$. Set point: $V_{\text{s}}$ = 2 mV, $I_{\text{t}}$ = 100 pA, $V_{\text{osc}}$ = 0.2 mV.} 
\end{figure}

The d$I$/d$V$ spectra of trilayer FeSe film reveals the same transition under an in-plane magnetic field. Trilayer FeSe films undergo a transition from a BCS state to a Sarma state when the Zeeman energy is larger than superconducting gap size as shown in Fig. \ref{TrilayerdIdV} (b-f), which resemble the spectrum transition features of bilayer FeSe films. The critical magnetic field of the Fermi level cross of the Zeeman splitting coherence peaks  exhibits strong anisotropy when the magnetic field is along different directions (Fig. \ref{TrilayerdIdV}).  

\clearpage

\section{Zeeman energy, zero bias conductance and gap size of trilayer FeSe film under in-plane magnetic fields}
\begin{figure}[htp]
	\begin{center}
		\includegraphics[width=6.5in]{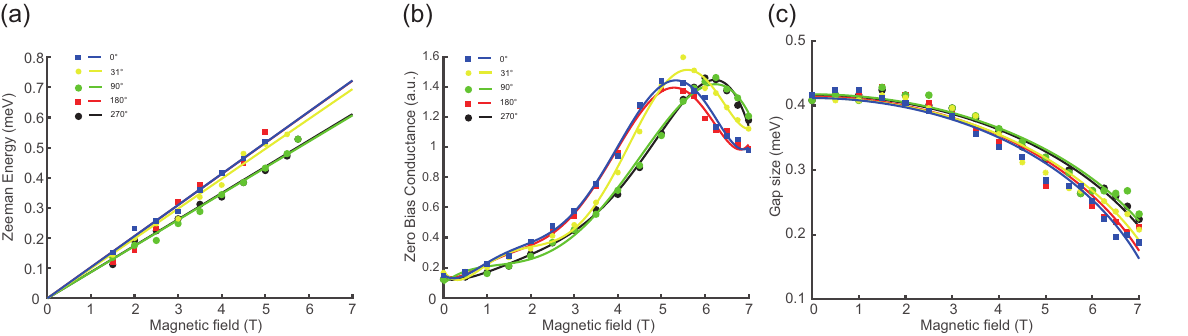}
	\end{center}
	\caption[]{\label{TrilayerCurve}(a)The Zeeman energy of trilayer FeSe under in-plane magnetic fields along five different directions. The solid circles are from experimental data, and the solid lines are the fitting curves. (b) The zero bias conductance of trilayer FeSe as a function of in-plane magnetic field along five different directions. (c) The superconducting gap size of trilayer FeSe as a function of in-plane magnetic fields along five different directions.
	} 
\end{figure}

The linear curves of the Zeeman energy as a function of the in-plane magnetic field are shown in Fig. \ref{TrilayerCurve}(a). The $g$ factor of trilayer FeSe film, extracted from those curves in Fig. \ref{TrilayerCurve}(a), shows strong in-plane anisotropy. The maximum $g$ factor is 1.78 when the field is along a$_{\text{Fe}}$ direction and the minimum is 1.49 when the field is along b$_{\text{Fe}}$ direction. The critical magnetic field extracted from ZBC peaks in Fig. \ref{TrilayerCurve}(b) displays the same anisotropy. As the field along a$_{\text{Fe}}$ direction, 5.3 T in-plane field induces the Fermi level cross of coherence peaks in trilayer FeSe film, while it needs 6.2 T as the magnetic field is along b$_{\text{Fe}}$. Moreover, the size of superconducting gap decreases when the in-plane field increases.

\clearpage

\section{The spatially resolved d$I$/d$V$ spectra of monolayer FeSe}

\begin{figure}[htp]
	\begin{center}
		\includegraphics[width=5in]{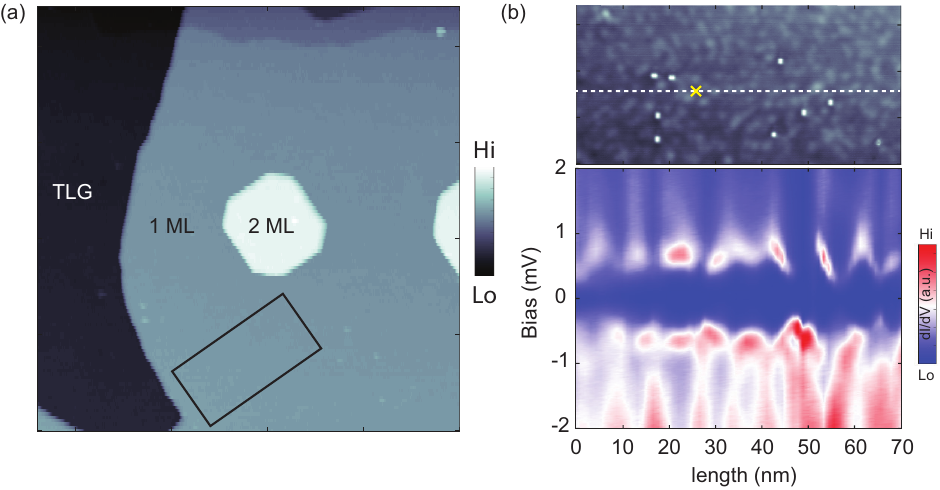}
	\end{center}
	\caption[]{\label{MLLinespectra}The spatially resolved d$I$/d$V$ spectra on a FeSe monolayer at zero magnetic field. (a) The topographic image of the FeSe thin films on trilayer graphene (TLG)/SiC(0001), consisting of FeSe monolayer and bilayer($V_{\rm{s}}$ = 3 V, $I_{\rm{t}}$ = 20 pA, 220 nm $\times$ 220 nm). (b) Upper panel: the topographic image of the FeSe monolayer marked by a black box in (a), illustrating a clean surface with a few Fe defects ($V_{\rm{s}}$ = 100 mV, $I_{\rm{t}}$ = 100 pA, 70 nm $\times$ 35 nm). Lower panel: Spatially resolved d$I$/d$V$ spectra acquired along the white dashed line shown in the upper panel, revealing modulation of the superconducting gap. The data presented in Figure 5 of the main text were recorded at the location marked by a yellow cross.
} 
\end{figure}
\clearpage

\section{The Zeeman splitting d$I$/d$V$ spectra of monolayer FeSe}

\begin{figure}[htp]
	\begin{center}
		\includegraphics[width=5in]{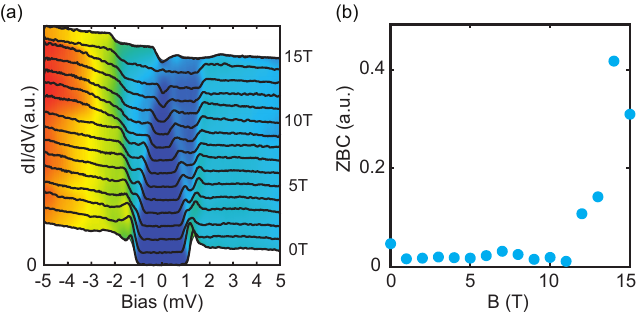}
	\end{center}
	\caption[]{\label{ZeemanSplitML} (a) An out-of-plane magnetic field-dependent d$I$/d$V$ spectra of the FeSe monolayer film with  $\Delta$ = 1.34 meV, measured from 0 T to 15 T ($V_{\rm{s}}$ = -10 mV, $I_{\rm{t}}$ = 100 pA, $V_{\rm{osc}}$ = 0.05 mV). (b) The magnetic field dependent zero-bias conductance extracted from (a).
}
\end{figure}
\clearpage


\end{document}